\documentclass[12pt,preprint]{aastex}

\shorttitle{\emph{Kepler}: oscillations in solar-type stars}
\shortauthors{Chaplin et al.}

\begin{document}

\title{Predicting the detectability of oscillations in solar-type
  stars observed by \emph{Kepler}}

\author{
   W.~J.~Chaplin\altaffilmark{1},
   H.~Kjeldsen\altaffilmark{2},
   T.~R.~Bedding\altaffilmark{3},
   J.~Christensen-Dalsgaard\altaffilmark{2},
   R.~L.~Gilliland\altaffilmark{4},
   S.~D.~Kawaler\altaffilmark{5},
   T.~Appourchaux\altaffilmark{6},
   Y.~Elsworth\altaffilmark{1},
   R.~A.~Garc\'ia\altaffilmark{7},
   G.~Houdek\altaffilmark{8},
   C.~Karoff\altaffilmark{2},
   T.~S.~Metcalfe\altaffilmark{9},
   J.~Molenda-\.Zakowicz\altaffilmark{10},
   M.~J.~P.~F.~G.~Monteiro\altaffilmark{11},
   M.~J.~Thompson\altaffilmark{9},
   G.~A.~Verner\altaffilmark{12,1},
   N.~Batalha\altaffilmark{13},
   W.~J.~Borucki\altaffilmark{13},
   T.~M.~Brown\altaffilmark{14},
   S.~T.~Bryson\altaffilmark{13},
   J.~L.~Christiansen\altaffilmark{15},
   B.~D.~Clarke\altaffilmark{15},
   J.~M.~Jenkins\altaffilmark{15},
   T.~C.~Klaus\altaffilmark{16},
   D.~Koch\altaffilmark{13},
   D.~An\altaffilmark{17},
   J.~Ballot\altaffilmark{18},
   S.~Basu\altaffilmark{19},
   O.~Benomar\altaffilmark{6},
   A.~Bonanno\altaffilmark{20},
   A.-M.~Broomhall\altaffilmark{1},
   T.~L.~Campante\altaffilmark{11,2},
   E.~Corsaro\altaffilmark{20},
   O.~L.~Creevey\altaffilmark{21,22},
   L.~Esch\altaffilmark{19},
   N.~Gai\altaffilmark{19,23},
   P.~Gaulme\altaffilmark{6},
   S.~J.~Hale\altaffilmark{1},
   R.~Handberg\altaffilmark{2},
   S.~Hekker\altaffilmark{1,24},
   D.~Huber\altaffilmark{3},
   S.~Mathur\altaffilmark{9},
   B.~Mosser\altaffilmark{25},
   R.~New\altaffilmark{26},
   M.~H.~Pinsonneault\altaffilmark{27},
   D.~Pricopi\altaffilmark{28},
   P.-O.~Quirion\altaffilmark{29},
   C.~R\'egulo\altaffilmark{30,22},
   I.~W.~Roxburgh\altaffilmark{12},
   D.~Salabert\altaffilmark{21,22},
   D.~Stello\altaffilmark{3},
   M.~D.~Suran\altaffilmark{28}
}
\altaffiltext{1}{School of Physics and Astronomy, University of Birmingham, Edgbaston, Birmingham, B15 2TT, UK}

\altaffiltext{2}{Department of Physics and Astronomy, Aarhus University, DK-8000 Aarhus C, Denmark}

\altaffiltext{3}{Sydney Institute for Astronomy (SIfA), School of Physics, University of Sydney, NSW 2006, Australia}

\altaffiltext{4}{Space Telescope Science Institute, Baltimore, MD 21218, USA}

\altaffiltext{5}{Department of Physics and Astronomy, Iowa State University, Ames, IA 50011, USA}

\altaffiltext{6}{Institut d'Astrophysique Spatiale, Universit\'e Paris XI -- CNRS (UMR8617), Batiment 121, 91405 Orsay Cedex, France}

\altaffiltext{7}{Laboratoire AIM, CEA/DSM -- CNRS -- Universit\'e Paris Diderot -- IRFU/SAp, 91191 Gif-sur-Yvette Cedex, France}

\altaffiltext{8}{Institute of Astronomy, University of Vienna, A-1180 Vienna, Austria}

\altaffiltext{9}{High Altitude Observatory and, Scientific Computing Division, National Center for Atmospheric Research, Boulder, Colorado 80307, USA}

\altaffiltext{10}{Astronomical Institute, University of Wroc\l{}aw, ul. Kopernika, 11, 51-622 Wroc\l{}aw, Poland}

\altaffiltext{11}{Centro de Astrof\'\i sica and Faculdade de Ci\^encias, Universidade do Porto, Rua das Estrelas, 4150-762, Portugal}

\altaffiltext{12}{Astronomy Unit, Queen Mary, University of London, Mile End Road, London, E1 4NS, UK}

\altaffiltext{13}{NASA Ames Research Center, MS 244-30, Moffett Field, CA 94035, USA}

\altaffiltext{14}{Las Cumbres Observatory Global Telescope, Goleta, CA 93117,USA}

\altaffiltext{15}{SETI Institute/NASA Ames Research Center, MS 244-30, Moffett Field, CA 94035, USA}

\altaffiltext{16}{Orbital Sciences Corporation/NASA Ames Research Center, Moffett Field, CA 94035, USA}

\altaffiltext{17}{Ewha Womans University, 11-1 Daehyun-Dong Seodaemun-Gu, Seoul 120-750 Korea}

\altaffiltext{18}{Institut de Recherche en Astrophysique et Plan\'etologie, Universit\'e de Toulouse, CNRS, 14 av E. Belin, 31400 Toulouse, France}

\altaffiltext{19}{Department of Astronomy, Yale University, P.O. Box 208101, New Haven, CT 06520-8101, USA}

\altaffiltext{20}{INAF Osservatorio Astrofisico di Catania, Via S.Sofia 78, 95123, Catania, Italy}

\altaffiltext{21}{Departamento de Astrof\'{\i}sica, Universidad de La Laguna, E-38206 La Laguna, Tenerife, Spain}

\altaffiltext{22}{Instituto de Astrof\'{\i}sica de Canarias, E-38200 La Laguna, Tenerife, Spain}

\altaffiltext{23}{Beijing Normal University, Beijing 100875, P.R. China}

\altaffiltext{24}{Astronomical Institute, "Anton Pannekoek",
University of Amsterdam, PO Box 94249, 1090 GE Amsterdam, The
Netherlands}

\altaffiltext{25}{LESIA, CNRS, Universit\'e Pierre et Marie Curie, Universit\'e Denis Diderot, Observatoire de Paris, 92195 Meudon cedex, France}

\altaffiltext{26}{Materials Engineering Research Institute, Faculty of
Arts, Computing, Engineering and Sciences, Sheffield Hallam
University, Sheffield, S1 1WB, UK}

\altaffiltext{27}{Department of Astronomy, The Ohio State University,
4055 McPherson Laboratory, 140 West 18th Avenue, Columbus, OH 43210,
USA}

\altaffiltext{28}{Astronomical Institute of the Romanian Academy,
Str. Cutitul de Argint, 5, RO 40557,Bucharest,RO}

\altaffiltext{29}{Canadian Space Agency, 6767 Boulevard de
l'A\'eroport, Saint-Hubert, QC, J3Y 8Y9, Canada}

\altaffiltext{30}{Departamento de Astrof\'{\i}sica, Universidad de La
Laguna, E-38206 La Laguna, Tenerife, Spain}

\begin{abstract}

Asteroseismology of solar-type stars has an important part to play in
the exoplanet program of the NASA \emph{Kepler Mission}. Precise and
accurate inferences on the stellar properties that are made possible
by the seismic data allow very tight constraints to be placed on the
exoplanetary systems. Here, we outline how to make an estimate of the
detectability of solar-like oscillations in any given \emph{Kepler}
target, using rough estimates of the temperature and radius, and the
\emph{Kepler} apparent magnitude.

\end{abstract}


\section{Introduction}
\label{sec:intro}

The primary objective of the NASA \emph{Kepler Mission} is to detect,
by the transit method, Earth-sized planets in the habitable zones of
solar-type main-sequence stars (Borucki et al. 2010; Koch et
al. 2010). Photometry of most of the stars is conducted at a long
cadence (LC) of 29.4\,minutes, but a subset of up to 512 stars can be
observed at a short cadence (SC) of 58.85\,s. The exquisite precision
and accuracy of the \emph{Kepler} data mean they are also well suited
to asteroseismic studies of stars (e.g., see Gilliland et al. 2010a,
Jenkins et al. 2010). The cadence of the SC data is rapid enough to
allow investigations of solar-like oscillations in solar-type stars,
where dominant periods are of the order of several minutes (Chaplin et
al. 2010, 2011a; Metcalfe et al. 2010).

Asteroseismology has an important role to play in the exoplanet
program, in that the precise and accurate inferences on the stellar
properties -- i.e., radius, mass and age -- that are made possible by
the seismic data allow very tight constraints to be placed on the
exoplanetary systems, e.g., tight estimation of the stellar radius
constrains the size of the planet (Kjeldsen et al. 2009; Stello et
al. 2009a; Christensen-Dalsgaard et al. 2010). Solar-type stars
flagged as potential planet hosts are therefore prime candidates for
asteroseismic study.

The LC lightcurves of solar-type stars that are flagged by the
Transiting Planet Search pipeline as having significant transit-like
features are modeled, and those stars having companion radii less than
two Jupiter radii are labeled as a Kepler Object of Interest (KOI)
(Batalha et al. 2010a). One of the factors that then determines
whether or not to assign the star to one of limited number of SC slots
-- allowing more rapid time-cadence observations -- is whether the
intrinsic properties and the apparent magnitude of the star would
allow the detection of solar-like oscillations. This paper describes
the straightforward procedure that has been implemented to answer this
question.

We use simple scaling relations to predict the oscillations and
granulation properties of any given KOI. The predictions use as input
basic information available on each star from the \emph{Kepler Input
Catalog} (or KIC; see, e.g., Batalha et al. 2010b). Together with the
known apparent magnitude of a given target, the scaling relation
predictions are used to construct an appropriate measure of the
signal-to-noise ratio in the frequency-power spectrum of the
solar-like oscillations, and, from that, we use some straightforward
statistics to estimate the probability of detection for the star.

To calibrate and verify the recipe we use asteroseismic results on
solar-type stars that were observed by \emph{Kepler} during the first
seven months of science operations. About 2000 stars, down to Kepler
apparent magnitude $Kp \simeq 12.5$, were selected as potential
solar-type targets based upon their KIC parameters. Each was observed
for one month at a time in SC. Timeseries were prepared for
asteroseismic analysis in the manner described by Garc\'ia et
al. (2011), and different teams applied their codes to attempt to
detect, and then extract the basic properties of, the solar-like
oscillations.  Several modeling teams also then applied codes to
estimate the stellar properties -- using the basic oscillations
parameters as inputs -- which we then also made use of in validation
of the recipe.

The layout of the rest of the paper is as follows. We introduce in
Section~\ref{sec:oscill} the basic principles underlying the detection
test. We then detail the scaling relation predictions that are
required for the test: Section~\ref{sec:scaling} gives the scaling
relations for the solar-like oscillations, while
Section~\ref{sec:bgtot} gives scaling relations for the granulation
and shot-noise contributions to the background power spectral
density. The various ingredients are brought together in
Section~\ref{sec:det}, in which the test is described in detail. The
predictions of the test are compared in Section~\ref{sec:res} to real
results returned on the solar-type ensemble that was observed for
asteroseismology. We conclude by presenting some general predictions
for detectability of \emph{Kepler} KOIs.

\section{Detection of solar-like oscillations}
\label{sec:oscill}

The frequency power spectra of solar-like oscillations in solar-type
stars present a pattern of peaks with near regular frequency
separations. The mode powers are modulated by an envelope. The
envelope has a bell-shaped appearance in many stars for which
solar-like oscillations have been observed (including the Sun).
Different techniques have been devised and applied to the
\emph{Kepler} SC data to detect signatures of solar-like oscillations
(e.g., see Chaplin et al. 2010; Campante et al. 2010; Hekker et
al. 2010; Huber et al. 2009; Karoff et al. 2010; Mosser \& Appourchaux
2009; Mathur et al. 2010; Roxburgh 2009). Some techniques rely on
extracting signatures of the near-regular frequency separations of the
oscillations; others search for signatures of the Gaussian-like power
excess due to the oscillations.

Here, we devise and apply a simple statistical detection test that is
based upon the second family of techniques. Our choice is predicated
on the requirement that the test should be straightforward to
implement on a large number of stars. Making an absolute prediction of
the detectability of the frequency separations is more complicated
than the test we adopt. Strictly speaking, it demands that the damping
rates of the solar-like oscillations, and the rotation and angle of
inclination of the star, are known. We therefore opt for a simpler
approach, which does not require assumptions to be made about these
properties. We do so in the knowledge that once the oscillations power
excess has been detected, the large frequency separation in solar-type
stars will also be readily extractable.

The detection test is based upon the total observed power due to
p-mode oscillations in the star. Let $P_{\rm tot}$ be the total
expected underlying mean power. As we shall show below, this is very
easily estimated from scaling relations. In the frequency-power
spectrum, the approximately Gaussian-shaped power envelope due to the
oscillations sits on top of slowly varying background power, which we
assume is dominated by contributions from shot/instrumental noise and
stellar granulation. When the observed power in the oscillations
relative to the background is high (i.e., high SNR), the power excess
due to the oscillations will be clearly visible. However, at low SNR,
statistical fluctuations in the background power may swamp the
oscillations signal, so that the excess is much harder to see.

If $B_{\rm tot}$ measures the total underlying background power across
the range in frequency occupied by the oscillations, then a ``global''
measure of SNR in the oscillations spectrum is given by
 \begin{equation}
 {\rm SNR_{tot}} = P_{\rm tot} / B_{\rm tot}.
 \label{eq:snr}
 \end{equation}
Provided ${\rm SNR_{tot}}$ is high enough, we may be able to claim an
unambiguous detection of the power excess due to the oscillations.  If
$N$ independent frequency bins have contributed to the estimation of
the SNR, we may then test it against $\chi^2$ $2N$-d.o.f. statistics
to determine if the SNR is sufficiently high that it would be hard to
explain by chance alone. This is a false-alarm approach.

Let us suppose that we demand that a false alarm test be passed at,
say, the 1\,\% level to merit a likely detection. We first estimate a
false-alarm threshold, ${\rm SNR_{thresh}}$, i.e., the SNR value above
which there is less than a 1 in 100 chance of a $\chi^2$
$2N$-d.o.f. distribution giving a normalized value greater than or
equal to this purely at random. We then use the scaling relations to
estimate the expected underlying ${\rm SNR_{tot}}$ of the target, and,
with that number in hand, we may then estimate the probability that an
observed value of the SNR for the star would be greater than, or equal
to, ${\rm SNR_{thresh}}$, i.e., that it would pass the false-alarm
test\footnote{If the underlying ${\rm SNR_{tot}}$ is greater than, or
equal to, ${\rm SNR_{thresh}}$ one might naively expect that the
chances of passing the false-alarm test would be 100\,\%. However, one
must remember that the real observations will be modified by the
$\chi^2$ $2N$-d.o.f. statistics, which will mean that not every case
would pass.}. This final probability serves as an estimate of the
detection probability of the star.\\

\section{Scaling relations for solar-like oscillations}
\label{sec:scaling}

\subsection{$\nu_{\rm max}$ and $\Delta\nu$}
\label{sec:nus}

The frequency at the peak of the power envelope of the oscillations,
where the observed modes present their strongest amplitudes, is
commonly referred to as $\nu_{\rm max}$.  Our ability to predict
$\nu_{\rm max}$ from the fundamental stellar properties allows us to
estimate where in frequency the oscillations would be most prominent.

We assume that $\nu_{\rm max}$ scales with the acoustic cut-off
frequency (Brown et al. 1991; Kjeldsen \& Bedding 1995) determined for
an isothermal stellar atmosphere (as verified observationally by
Mosser et al. 2010). This scaling has been shown to provide good
predictions for observed solar-like oscillation spectra. This gives,
in terms of the fundamental stellar properties, the following scaling
relation:
 \begin{equation}
 \nu_{\rm max} = \nu_{\rm max,\odot}\,
 \left( \frac{M}{\rm M_{\odot}} \right)
 \left( \frac{R}{\rm R_{\odot}} \right)^{-2}
 \left( \frac{T_{\rm eff}}{{\rm T_{eff,\odot}}}\right)^{-0.5},
 \label{eq:numax}
 \end{equation}
where we have scaled against the solar values of $\nu_{\rm max,\odot}
= 3150\,\rm \mu Hz$ and ${\rm T_{eff,\odot}} = 5777\,\rm K$.

The most striking characteristic frequency separations in the
oscillation spectra are the so-called large frequency separations,
$\Delta\nu$. These are the separations between consecutive overtones,
$n$, having the same spherical angular degree, $l$. The average value
of the large frequency separation scales very well with the square
root of the mean density of the star (e.g., see Christensen-Dalsgaard
1993). Here, we use this scaling and calibrate estimated values
against the observed average large separation of the Sun, i.e.,
 \begin{equation}
 \Delta\nu = \Delta\nu_{\odot}
 \left( \frac{M}{\rm M_{\odot}} \right)^{0.5}
 \left( \frac{R}{\rm R_{\odot}} \right)^{-1.5},
 \label{eq:dnu}
 \end{equation}
with $\Delta\nu_{\odot} = 134.9\,{\rm \mu Hz}$. It has also recently
been shown (Stello et al. 2009b) that solar-like oscillations in
main-sequence stars follow to good approximation the following
relation:
 \begin{equation}
 \Delta\nu = \Delta\nu_{\odot} \left( \frac{\nu_{\rm max}}{\nu_{\rm
 max,\odot}} \right)^{0.75}.
 \label{eq:dnunumax}
 \end{equation}
Use of Equation~\ref{eq:dnunumax} allows us to eliminate $M$ from
Equations~\ref{eq:numax} and~\ref{eq:dnu}, to yield scaling relations
that depend only on $T_{\rm eff}$ and $R$. The process of elimination
yields the scaling relation $M \propto T_{\rm eff}^{1.5}$, which is
close to the relation $M \propto T_{\rm eff}^{1.7}$ in Noyes et
al. (1984). The scaling relations for $\Delta\nu$ and $\nu_{\rm max}$
in $R$ and $T_{\rm eff}$ are:
 \begin{equation}
 \nu_{\rm max} = \nu_{\rm max,\odot}\,
 \left( \frac{R}{\rm R_{\odot}} \right)^{-2}
 \left( \frac{T_{\rm eff}}{{\rm T_{eff,\odot}}}\right),
 \label{eq:numax1}
 \end{equation}
 \begin{equation}
 \Delta\nu = \Delta\nu_{\odot}\,
 \left( \frac{R}{\rm R_{\odot}} \right)^{-1.5}
 \left( \frac{T_{\rm eff}}{{\rm T_{eff,\odot}}}\right)^{0.75}.
 \label{eq:dnu1}
 \end{equation}

\subsection{Mode amplitudes}
\label{sec:amp}

In order to comment on mode detectability, we must be able to predict
the mode amplitudes. We base our predictions on Kjeldsen \& Bedding
(1995) and Samadi et al. (2007). We make predictions for observations
made in intensity (relevant to \emph{Kepler} photometry). The prediction is
that the maximum oscillation intensity amplitudes $A_{\rm max}$ of the
radial ($l=0$) modes scale like
 \begin{equation}
 A_{\rm max} = A_{\rm max,\odot}~\beta\,
 \left( \frac{L/{\rm L_{\odot}}}{M/{\rm M_{\odot}}} \right)^{s}
 \left( \frac{T_{\rm eff}}{{\rm T_{eff,\odot}}} \right)^{-2},
 \label{eq:amp}
 \end{equation}
with the coefficient $s$ often quoted between 0.7 and 1.0. We follow
Kjeldsen \& Bedding (1995), and assume that the temperature dependence
is quadratic.  We have also included another coefficient, $\beta$,
because without any further correction the above relation is known to
overestimate the amplitudes in the hottest solar-type stars (e.g., see
Houdek 2006, and references therein). To estimate the solar
\textsc{rms} value $A_{\rm max,\odot}$ we used the \emph{Kepler}
response function in wavelength, and known measurements of the solar
amplitudes made in narrow-band observations by the three wavelength
channels of the VIRGO/SPM instrument onboard SoHO (see Ballot, in
preparation). From the SPM data we defined the observed solar
amplitude as a function of wavelength, and then weighted this curve by
the \emph{Kepler} response function to give an estimated \textsc{rms}
value of $A_{\rm max,\odot} \sim 2.5\,\rm ppm$.


\begin{figure*}
\epsscale{1.0}
\plottwo{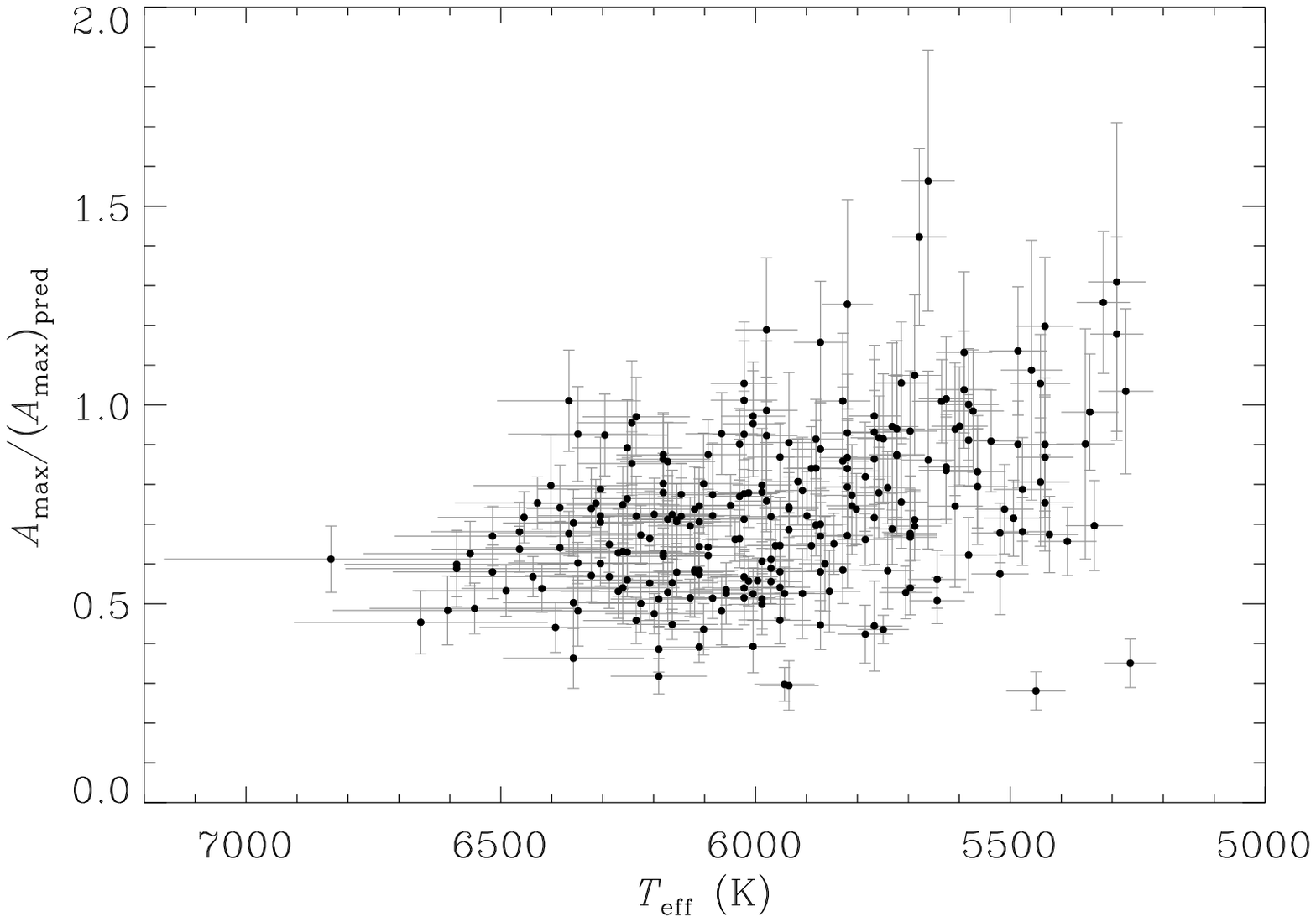}{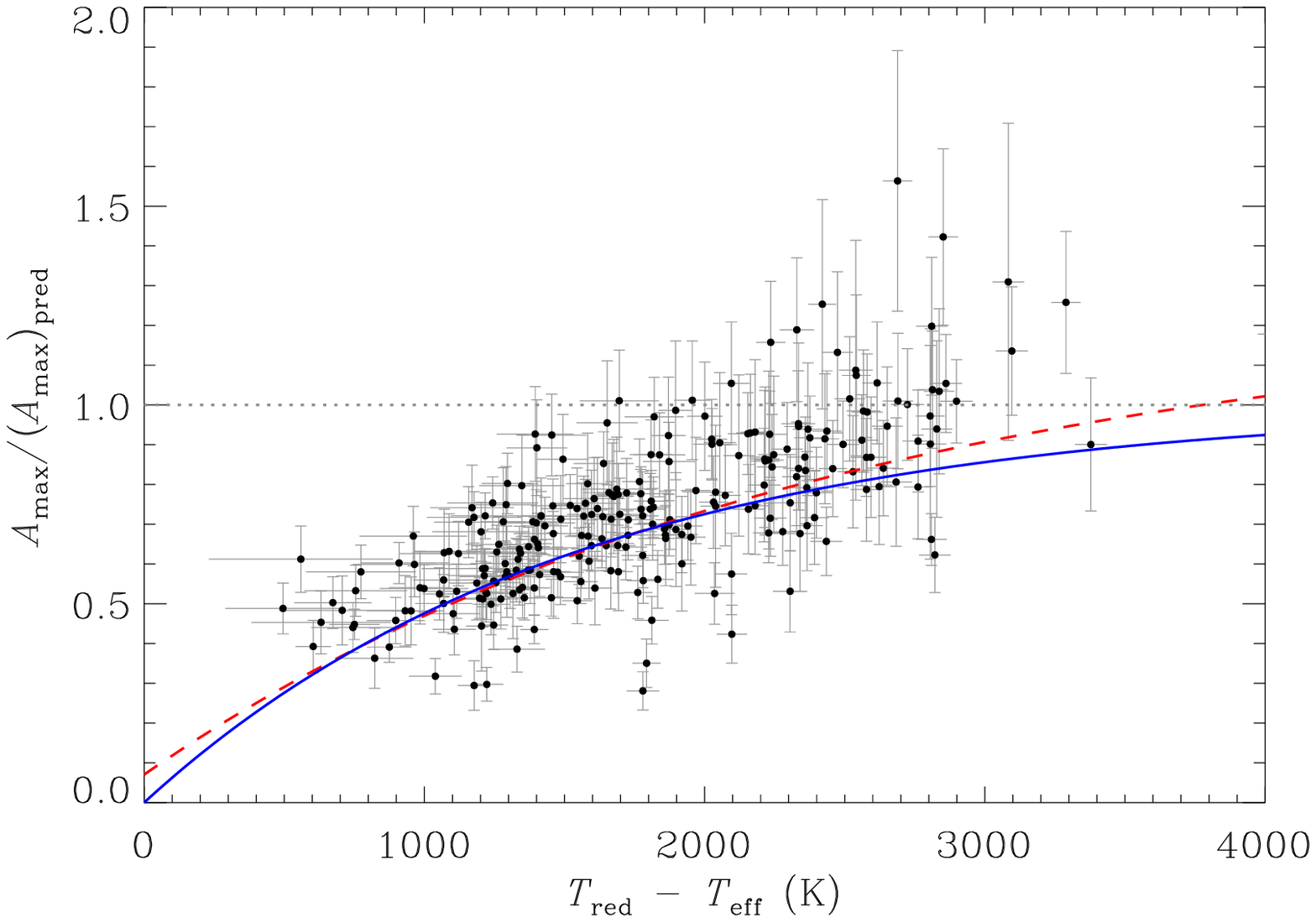}
\caption{Plots of ratio of the observed to the predicted maximum mode
 amplitudes, assuming that the predicted amplitudes scale like
 $(L/M)^{1.0}$.}
\label{fig:beta}
\end{figure*}


We have used results of \emph{Kepler} observations of solar-type stars
to calibrate the size of the correction $\beta$. Results on
$\Delta\nu$ and $\nu_{\rm max}$ were used by modeling teams, together
with $T_{\rm eff}$, to provide estimates of the stellar radii and
masses (e.g., see Stello et al. 2009a; Basu et al. 2010; Quirion et
al. 2010; Gai et al. 2010). These estimated stellar properties were
used to calculate predicted maximum amplitudes using
Equation~\ref{eq:amp}, with the unknown (at this stage) $\beta$ set to
unity. We then calculated the ratios of the observed amplitudes,
$A_{\rm max}$ (extracted by the Octave pipeline), to the predicted
amplitudes, $(A_{\rm max})_{\rm pred}$, for different assumed values
of $s$. We now go on to discuss the results obtained for $s=1.0$.

The ratios for $s=1.0$ are plotted in Fig.~\ref{fig:beta} against
$T_{\rm eff}$ (left-hand panels) and $T_{\rm red} - T_{\rm eff}$
(right-hand panels). We define $T_{\rm red}$ to be the temperature on
the red edge of the radial-mode $\delta$-Scuti instability strip,
given in Houdek et al. (1999), at the luminosity, $L$, of a given
star:
 \begin{equation}
 T_{\rm red} = T_{\rm red,\odot} (L/{\rm L_{\odot}})^{-0.093}
 \label{eq:teffred}
 \end{equation}
with $T_{\rm red,\odot}=8907\,\rm K$. Both panels of
Fig.~\ref{fig:beta} show a decrease in the ratio of the observed to
predicted amplitudes as $T_{\rm eff}$ increases.  A more clearly
defined trend is present in the $T_{\rm red} - T_{\rm eff}$ plot, and
we therefore chose to use fits performed on these data to fix the
correction $\beta$. We fitted the following functions: first
 \begin{equation}
 \beta = 1 - \exp \left( - \frac{T_{\rm red} - T_{\rm eff}}{\Delta T}
 \right),
 \label{eq:beta2}
 \end{equation}
and then the less restrictive
 \begin{equation}
 \beta = c_0 - c_1 \exp \left( - \frac{T_{\rm red} - T_{\rm eff}}{\Delta T}
 \right)
 \label{eq:beta1}
 \end{equation}
where $c_0$, $c_1$ and $\Delta T$ were the free parameters to be
optimized, and weights for the fits were fixed by the uncertainties on
the amplitude ratios. The correction seeks to allow for the
attenuation of the amplitudes at high $T_{\rm eff}$ (which maps to low
$T_{\rm red} - T_{\rm eff}$), and since at progressively lower $T_{\rm
eff}$ one expects the amplitude predictions to be much more robust, we
require a function that turns over at high $T_{\rm red} - T_{\rm eff}$
(hence the choice of an exponential).  The solid blue and dashed red
lines in the right-hand panel of Fig.~\ref{fig:beta} show the
best-fitting Equations~\ref{eq:beta2} and~\ref{eq:beta1},
respectively. The best-fitting $\chi^2$ for the two models were almost
indistinguishable (about 2.3), and as such we chose to adopt the
simpler Equation~\ref{eq:beta2} as our model for $\beta$, with its
best-fitting $\Delta T = 1550\,\rm K$. A fit to a simple straight-line
model gave a marginally better $\chi^2$ than either of the two
exponential models. The straight-line model does of course
fail to capture the requirement that the amplitude ratio curve turn
over at higher $T_{\rm red} - T_{\rm eff}$, and its use would lead to
the prediction of unphysically high amplitudes in the coolest
solar-type stars.

We also performed the above analyses assuming the predicted
amplitudes scaled like $(L/M)^{0.7}$ (i.e., with $s=0.7$ in
Equation~\ref{eq:amp}). However, we found that we had to increase
arbitrarily our assumed value of $A_{\rm max,\odot}$ -- from 2.5 to
3.1\,ppm -- otherwise most of the amplitude ratios lay significantly
above unity. We therefore opted to fix $s=1.0$ (as per the
results outlined above).

To summarize, we have, using the scaling relations in $R$ and $T_{\rm
eff}$, and taking $s=1.0$,
 \begin{equation}
 A_{\rm max} = 2.5\,\beta\,
 \left( \frac{R}{\rm R_{\odot}} \right)^{2}
 \left( \frac{T_{\rm eff}}{{\rm T_{eff,\odot}}}\right)^{0.5},
 \label{eq:amp1}
 \end{equation}
with $\beta$ calibrated by Equation~\ref{eq:beta2} using $\Delta T =
1550\,\rm K$.

 \section{Scaling relations for background power spectral density}
 \label{sec:bgtot}

We assume that there are two dominant contributions to the background
power spectral density in the vicinity of $\nu_{\rm max}$ in the
frequency-power spectrum. The largest contribution comes from
instrumental/shot noise, which we define in terms of a simple offset
$b_{\rm instr}$.  When observations of the oscillations are made in
intensity, stellar granulation also contributes significant power
which must be accounted for in estimation of the total background.  We
model the granulation power as a Lorentzian function centred on zero
frequency (e.g., Harvey 1985), with a characteristic power spectral
density, $b_{\rm gran}$, and a characteristic timescale, $\tau_{\rm
gran}$. The power spectral density $P_{\rm gran}$ at frequency
$\nu_{\rm max}$ is then given by:
 \begin{equation}
 P_{\rm gran} = \frac{b_{\rm gran}}{1+(2\pi \tau_{\rm gran} \nu_{\rm
 max})^2}\,\rm~ppm^2\,\mu Hz^{-1}.
 \label{eq:harvey}
 \end{equation}
We now go on to describe scaling relations for each of the free
parameters needed to model the total background, beginning with those
describing the granulation term.

 \subsection{Scaling relations for stellar granulation}
 \label{sec:scalegran}

We follow Huber et al. (2009) in assuming that $\tau_{\rm gran}$
scales inversely with $\rm \nu_{\rm max}$, as proposed by Kjeldsen \&
Bedding (in preparation). This means that the denominator of
Equation~\ref{eq:harvey} will be constant for all stars. Taking solar
values of $\tau_{\rm gran,\odot} = 210\,\rm sec$ and $\nu_{\rm
max,\odot} = 3150\,\rm \mu Hz$, we have that:
 \begin{equation}
 P_{\rm gran} \approx 0.055 b_{\rm gran}\,\rm~ppm^2\,\mu Hz^{-1}.
 \end{equation}
To describe $b_{\rm gran}$, we assume that the typical size of a
convective granule is proportional to the scale height for an
isothermal atmosphere, and that all granules behave in a statistically
independent manner so that the total \textsc{rms} fluctuation scales
inversely as the square root of the number of observed granules. This
leads to (e.g., Schwarzschild 1975; Kjeldsen \& Bedding, in
preparation; see also Baudin et al. 2006):
 \begin{equation}
 b_{\rm gran} \propto
 \left( \frac{L}{\rm L_{\odot}} \right)^{2}
 \left( \frac{M}{\rm M_{\odot}} \right)^{-3}
 \left( \frac{T_{\rm eff}}{T_{\rm eff,\odot}} \right)^{-5.5}.
 \end{equation}
With $M \propto T_{\rm eff}^{1.5}$ (see Section~\ref{sec:nus}), the
above implies that
 \[
 P_{\rm gran} \propto b_{\rm gran} \propto
 \left( \frac{R}{\rm R_{\odot}} \right)^{4}
 \left( \frac{T_{\rm eff}}{{\rm T_{eff,\odot}}}\right)^{-2} 
 \]
 \begin{equation}
 ~~~~~~~~~~\propto \left( \frac{\nu_{\rm max}}{\nu_{\rm max,\odot}} \right)^{-2}.
 \label{eq:pg1}
 \end{equation}


\begin{figure*}
\epsscale{1.0}
\plottwo{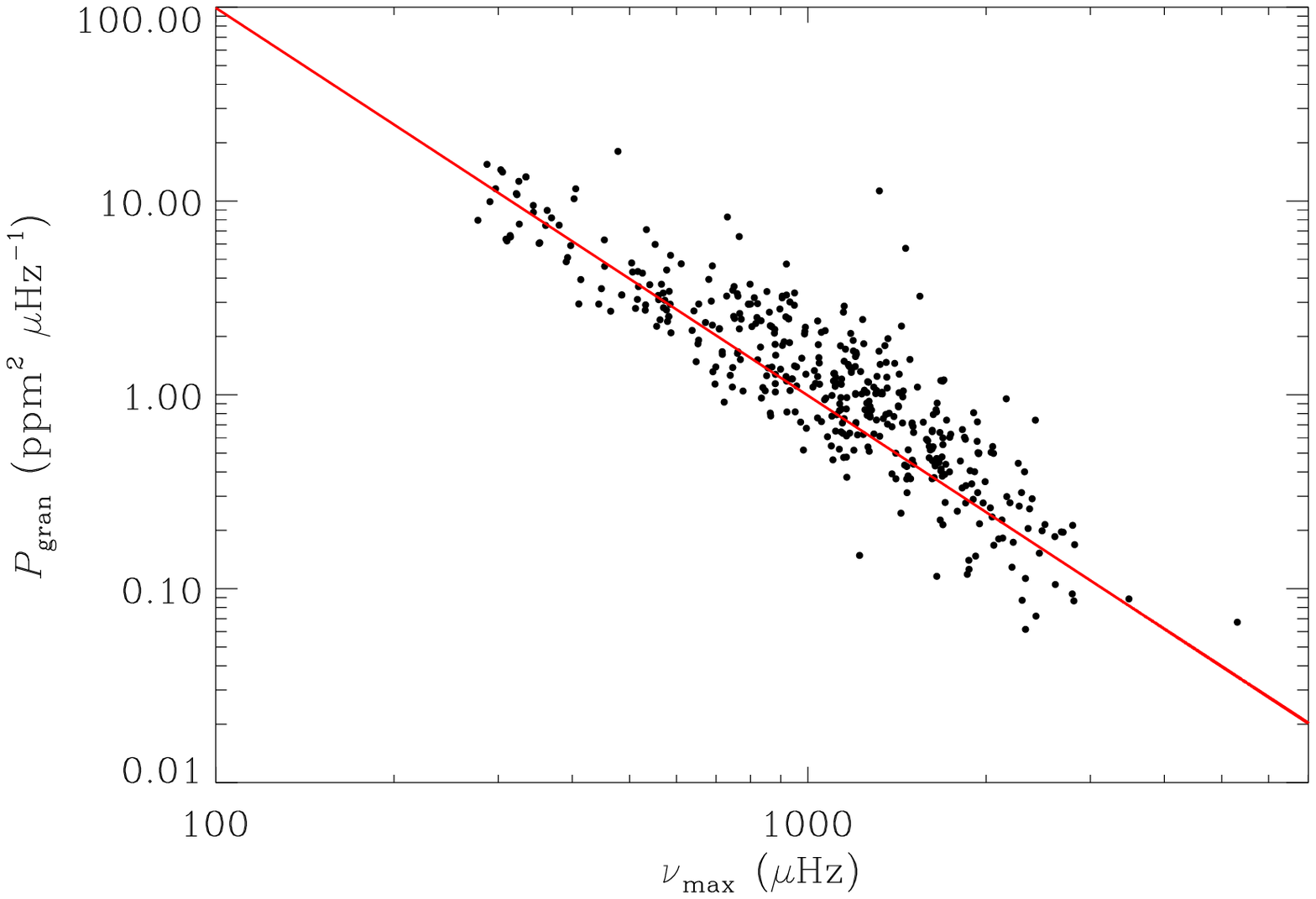}{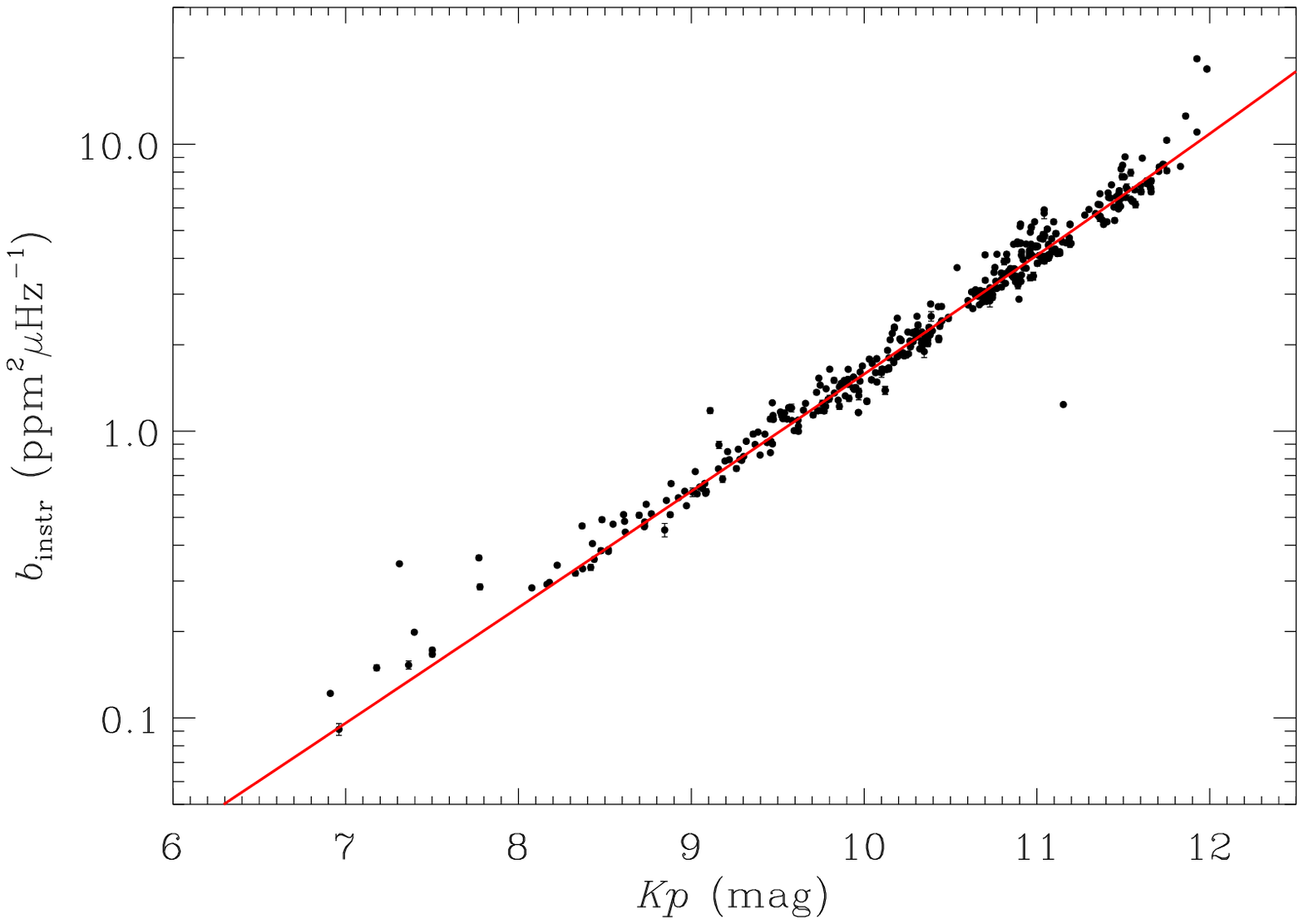}
\caption{Left-hand panel: Best-fitting $P_{\rm gran}$ of solar-type
stars observed by \emph{Kepler} that showed oscillations. The line
shows the prediction from Equation~\ref{eq:bmax}. Right-hand panel:
Best-fitting $b_{\rm instr}$ of the same stars. The line shows the
prediction from the minimal-term noise model in Gilliland et
al. (2010b).}
\label{fig:off}
\end{figure*}


We have tested Equation~\ref{eq:pg1} using results from the
\emph{Kepler} survey on solar-type stars. The left-hand panel of
Fig.~\ref{fig:off} plots the best-fitting $P_{\rm gran}$ of solar-type
stars with detected oscillations (given by a fit of a Harvey-like
model, plus a flat offset, to the background power spectral
density). The best-fitting power law is found to be $P_{\rm gran}
\propto \nu_{\rm max}^{-1.8}$, which is close to the prediction
above. Using appropriate normalization from the real data, we
therefore adopt:
 \begin{equation}
 P_{\rm gran} \approx 0.1 \left( \frac{\nu_{\rm max}}{\nu_{\rm
 max,\odot}} \right)^{-2}\,\rm ppm^2\,\mu Hz^{-1}.
 \label{eq:bmax}
 \end{equation}
The red line in Fig.~\ref{fig:off} shows the prediction from
Equation~\ref{eq:bmax}.

 \subsection{Scaling relations for instrumental/shot noise}
 \label{sec:scaleshot}

We use the ``minimal term'' noise model for \emph{Kepler} presented in
Gilliland et al. (2010b) in order to predict the \textsc{rms} noise,
$\sigma$, per 58.85-sec integration in the time domain:
 \begin{equation}
 \sigma = \frac{10^6}{c} \left( c+9.5 \times 10^5(14/Kp)^5
 \right)^{1/2}\,\rm ppm
 \label{eq:noise}
 \end{equation}
where $c = 1.28 \times 10^{0.4(12-Kp)+7}$ detections~per~cadence. The
expected noise power spectral density in the frequency domain due to
the instrumental noise is then given approximately by:
 \begin{equation}
 b_{\rm instr} = 2 \times 10^{-6}\,\sigma^2\,\Delta t~\,\rm ppm^2\,\mu Hz^{-1},
 \end{equation}
where $\Delta t$ is the mean 58.85-sec cadence.

The right-hand panel of Fig.~\ref{fig:off} plots the best-fitting
$b_{\rm instr}$ of solar-type stars showing detected oscillations. The
solid line shows the prediction from Equation~\ref{eq:noise}. We see
acceptable agreement between the observations and the predictions.

\section{The detection test}
\label{sec:det}

\subsection{Estimation of ${\rm SNR_{tot}}$}
\label{sec:snrtot}

First, we must compute the expected underlying global SNR of the
target of interest, ${\rm SNR_{tot}}$.

As noted previously, we assume that the power in the oscillations is
modulated by a Gaussian-shaped envelope in frequency. We wish to
calculate the total expected power underneath the envelope. The FWHM
of the power envelope is assumed to be equal to $\nu_{\rm max}/2$
(e.g., see Stello et al. 2007, and references therein; also Mosser et
al. 2010), implying that most of the total mode power is contained
within a range $\pm \nu_{\rm max}/2$ around $\nu_{\rm max}$.

Each segment $\Delta\nu$ of the spectrum will contain power due to the
visible $l=0$, 1, 2 and 3 modes (although note that the relative
contribution from the $l=3$ modes is extremely weak). The $l=0$ mode
at the centre of the Gaussian-shaped envelope will have a power
approximately equal to $A_{\rm max}^2$. The neighbouring $l=1$, 2 and
3 modes in the central $\Delta\nu$ will contribute a further $\approx
2.1A_{\rm max}^2$ of power (e.g., see Ballot, in preparation; Kjeldsen
et al. 2008). This relative contribution is fixed by the relative
visibilities of the different $l$. The total oscillations power in the
central segment is therefore equal to $\approx 3.1A_{\rm max}^2$.

\begin{figure}
\epsscale{1.0}
\plotone{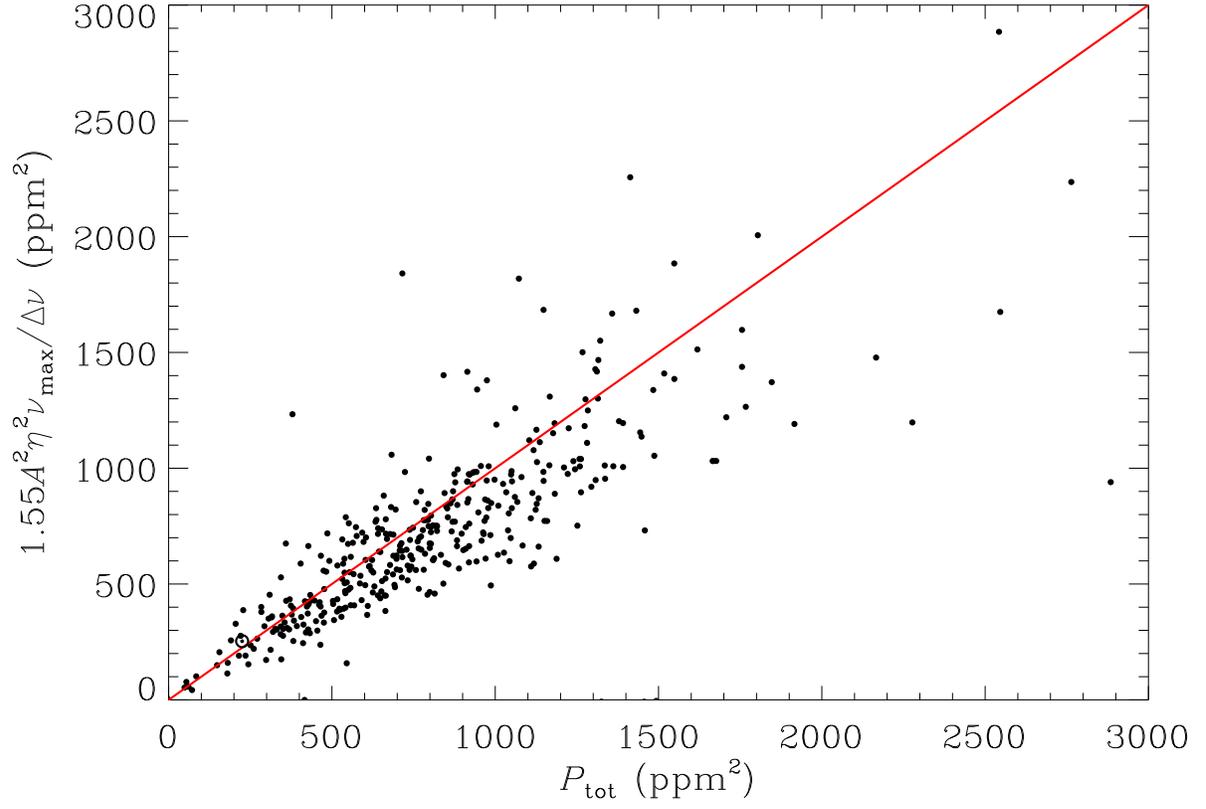}

\caption{$P_{\rm tot}$ constructed with Equation~\ref{eq:ptot1a} from
the observed $A_{\rm max}$, $\nu_{\rm max}$ and $\Delta\nu$ of the
stars showing detected oscillations against the actual measured
$P_{\rm tot}$ given by the sum of the power excess above the fitted
background of each frequency-power spectrum.}

\label{fig:ptot}
\end{figure}

To obtain the total mean power, $P_{\rm tot}$, underneath the
Gaussian-shaped envelope we must then add the contributions from all
other segments of $\Delta\nu$ that fall in the range $\pm \nu_{\rm
max}/2$ around $\nu_{\rm max}$. In all, $\nu_{\rm max} / \Delta\nu$
segments will contribute, with the average power of the contributing
segments being $\sim 0.5$ times that of the central segment (the
factor of 0.5 following from the fall-off in frequency of the Gaussian
function). This implies that:
 \begin{equation}
 P_{\rm tot} \approx 1.55 A_{\rm max}^2 \frac{\nu_{\rm max}}{\Delta\nu}.
 \label{eq:ptot}
\end{equation}
We must also make a correction for the fact that the high fractional
duty cycle of the \emph{Kepler} integrations, coupled with a cadence
of nearly one minute, leads to apodization of the oscillations signal
the closer $\nu_{\rm max}$ is to the Nyquist frequency ($\sim
8496\,\rm \mu Hz$) of the SC sampling. The amplitudes are suppressed
by the factor
 \begin{equation}
 \eta = {\rm sinc} \left[ \pi/2 \left( \frac{\nu_{\rm max}}{8496\,\rm \mu
 Hz}\right) \right],
 \end{equation}
so that Equation~\ref{eq:ptot} should actually be written as
 \begin{equation}
 P_{\rm tot} \approx 1.55 A_{\rm max}^2\,\eta^2\,\frac{\nu_{\rm max}}{\Delta\nu}.
 \label{eq:ptot1a}
\end{equation}
Fig.~\ref{fig:ptot} offers confirmation that Equation~\ref{eq:ptot1a}
does indeed provide a reasonably good description of $P_{\rm tot}$,
thereby implying that the Gaussian-envelope description provides a good
model for the total observed oscillations power. It shows $P_{\rm
tot}$, as constructed with Equation~\ref{eq:ptot1a} from the observed
$A_{\rm max}$, $\nu_{\rm max}$ and $\Delta\nu$ of stars showing
detected oscillations, against their actual measured $P_{\rm tot}$, as
given by the sum of the power excess above the fitted background in
each frequency-power spectrum. The red line shows the one-to-one
relation, which the data are seen to follow to good approximation.

We take one additional step for the detection recipe, and use
Equations~\ref{eq:numax1},~\ref{eq:dnu1} and~\ref{eq:amp1} to
re-express $\nu_{\rm max}$, $\Delta\nu$ and $A_{\rm max}$ in terms of
$R$ and $T_{\rm eff}$, to give the following relation:
 \begin{equation}
 P_{\rm tot} \approx 225\,{\rm ppm^2}\,\beta^2\,\eta^2\,
 \left( \frac{R}{\rm R_{\odot}} \right)^{3.5}
 \left( \frac{T_{\rm eff}}{{\rm T_{eff,\odot}}}\right)^{1.25}.
 \label{eq:ptot1}
\end{equation}


Turning next to the background, the total underlying background power
in the frequency-power spectrum across the range $\nu_{\rm max}$ is
given approximately by:
 \begin{equation}
 B_{\rm tot} \approx b_{\rm max} \nu_{\rm max},
 \label{eq:bgtot}
 \end{equation}
where $b_{\rm max}$ is the background power spectral density at the
frequency $\nu_{\rm max}$, i.e.,
 \begin{equation}
 b_{\rm max} = b_{\rm instr} + P_{\rm gran}.
 \end{equation}
For simplicity, we have ignored the frequency dependence of the
background, assuming instead that the value $b_{\rm max}$ captures the
average value of the background to sufficient accuracy.  We may then
calculate the global SNR in the total oscillations power as ${\rm
SNR_{tot}} = P_{\rm tot} / B_{\rm tot}$ (Equation~\ref{eq:snr}).

\subsection{Estimation of detection probability}
\label{sec:snrthr}

A total of $N$ independent frequency bins will contribute to
estimation of the underlying $P_{\rm tot}$ and $B_{\rm tot}$, and
hence $\rm SNR_{tot}$. We may therefore test $\rm SNR_{tot}$ against
$\chi^2$ $2N$-d.o.f. statistics to determine if the observed SNR would
be large enough that it would be hard to explain by chance alone. If
$T$ is the length of the timeseries, then $N$ will be given by:
 \begin{equation}
 N = \left( \frac{\nu_{\rm max}/{\rm \mu Hz} \times T/{\rm sec}}{10^6}
 \right).
 \label{eq:N}
 \end{equation}
The probability of obtaining any value ${\rm SNR}'$ above a given
level ${\rm SNR}$ is given by:
 \begin{equation}
 P({\rm SNR}' \ge {\rm SNR}, N) = \int_{x}^{\infty} 
  \frac{\exp{(-x')}}{\Gamma(N)} x'^{(N-1)}\,{\rm d}x',
 \label{eq:prob}
 \end{equation}
where $x=1+\rm SNR$ and $\Gamma$ is the Gamma function. Note the
normalization of $x$: as ${\rm SNR}$ tends to zero (i.e., no
oscillation signal present), we demand that $x$ tends to unity.

To flag a possible detection, we would demand that the \emph{observed}
SNR of the star exceed some SNR threshold, ${\rm SNR_{thresh}}$,
corresponding to a fractional false-alarm probability, $p$ (e.g.,
$p=0.01$, or 1\,\%). The required false-alarm threshold follows by
solving:
 \begin{equation}
 P({\rm SNR}' \ge {\rm SNR_{thresh}}, N) = p.
 \end{equation}
The probability that the observed SNR would exceed ${\rm
SNR_{thresh}}$ is then given by
 \begin{equation}
 P_{\rm final} = \int_{y}^{\infty} 
  \frac{\exp{(-y')}}{\Gamma(N)} y'^{(N-1)}\,{\rm d}y',
 \end{equation}
with
\begin{equation}
y = (1+{\rm SNR_{thresh}})/(1+{\rm SNR_{tot}}).
\end{equation}
We may regard $P_{\rm final}$ as providing an approximate estimate of
the probability of detecting solar-like oscillations in the star.  The
equation for $N$ captures the sensitivity of the detection probability
to the length of the observations, $T$. As $T$, and hence $N$,
increases, so the relative statistical fluctuations in $B_{\rm tot}$
will decrease in magnitude; hence, for a given underlying ${\rm
SNR_{tot}}$, the excess power due to the oscillations will be more
clearly visible against the background (as reflected in a reduction in
the size of ${\rm SNR_{thresh}}$).

\section{Results}
\label{sec:res}

We now apply our detection test to the solar-type stars observed for
one month each by \emph{Kepler} during the first seven months of
science operations. We use the KIC $R$ and $T_{\rm eff}$ and apparent
magnitude of each target to predict whether oscillations would be
detectable in one month. We then compare our predictions, in a
statistical sense, with the actual results from \emph{Kepler}. We only
consider stars having an observed, or predicted, $\nu_{\rm max} \ge
350\,\rm \mu Hz$, which takes us approximately to the base of the
red-giant branch.


\begin{figure*}
\epsscale{0.45}
\plotone{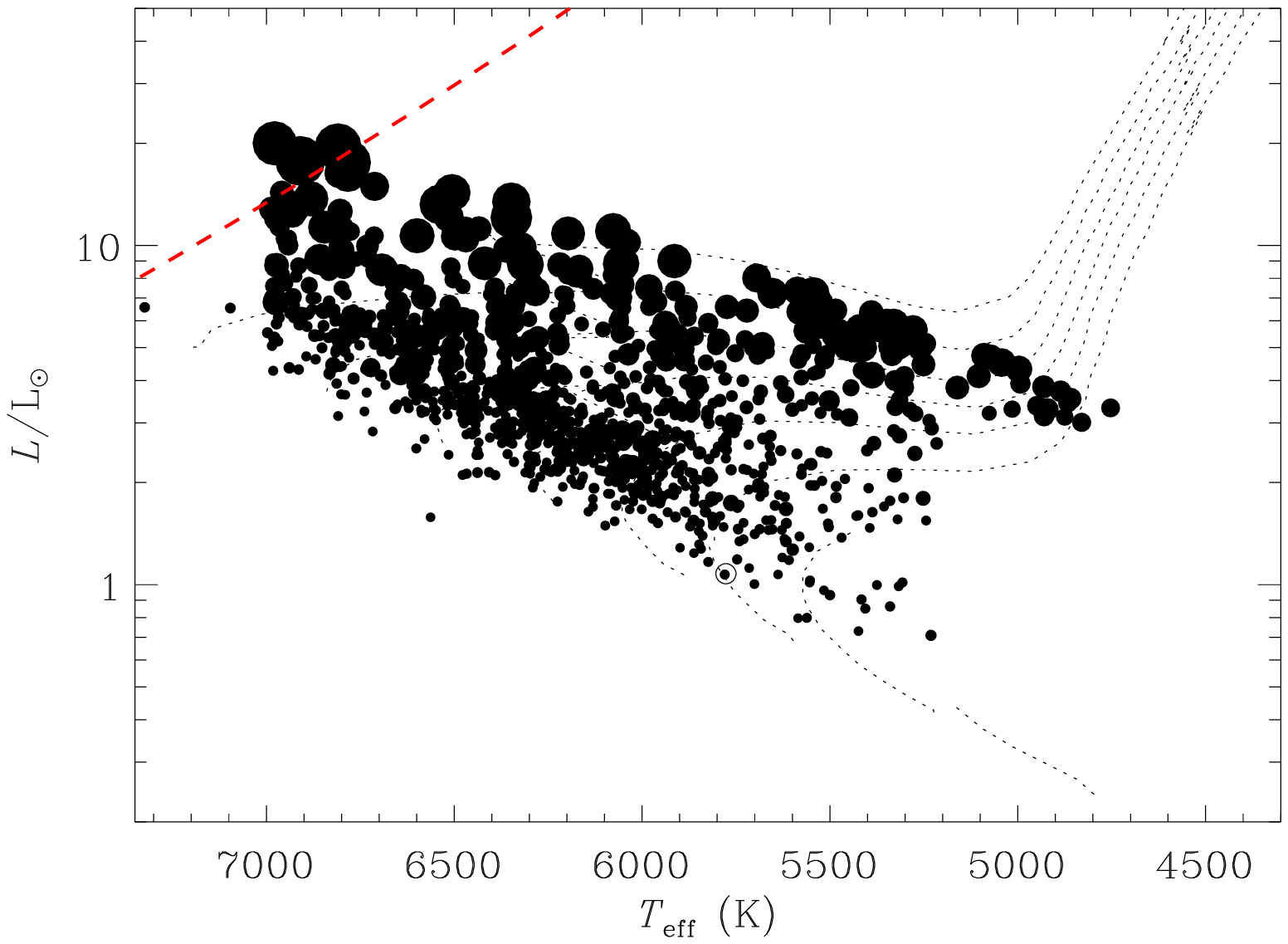}
\epsscale{1.0}
\plottwo{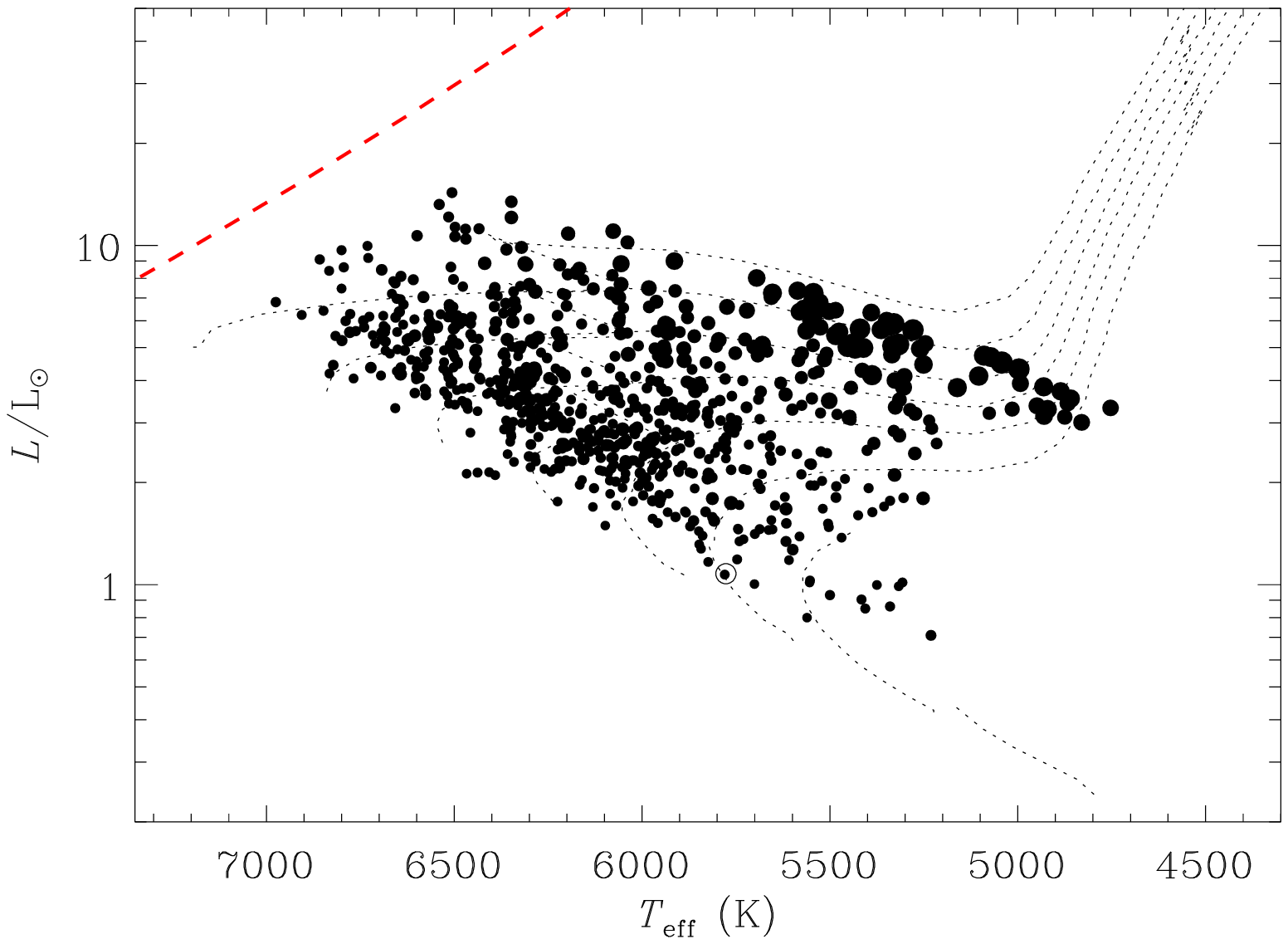}{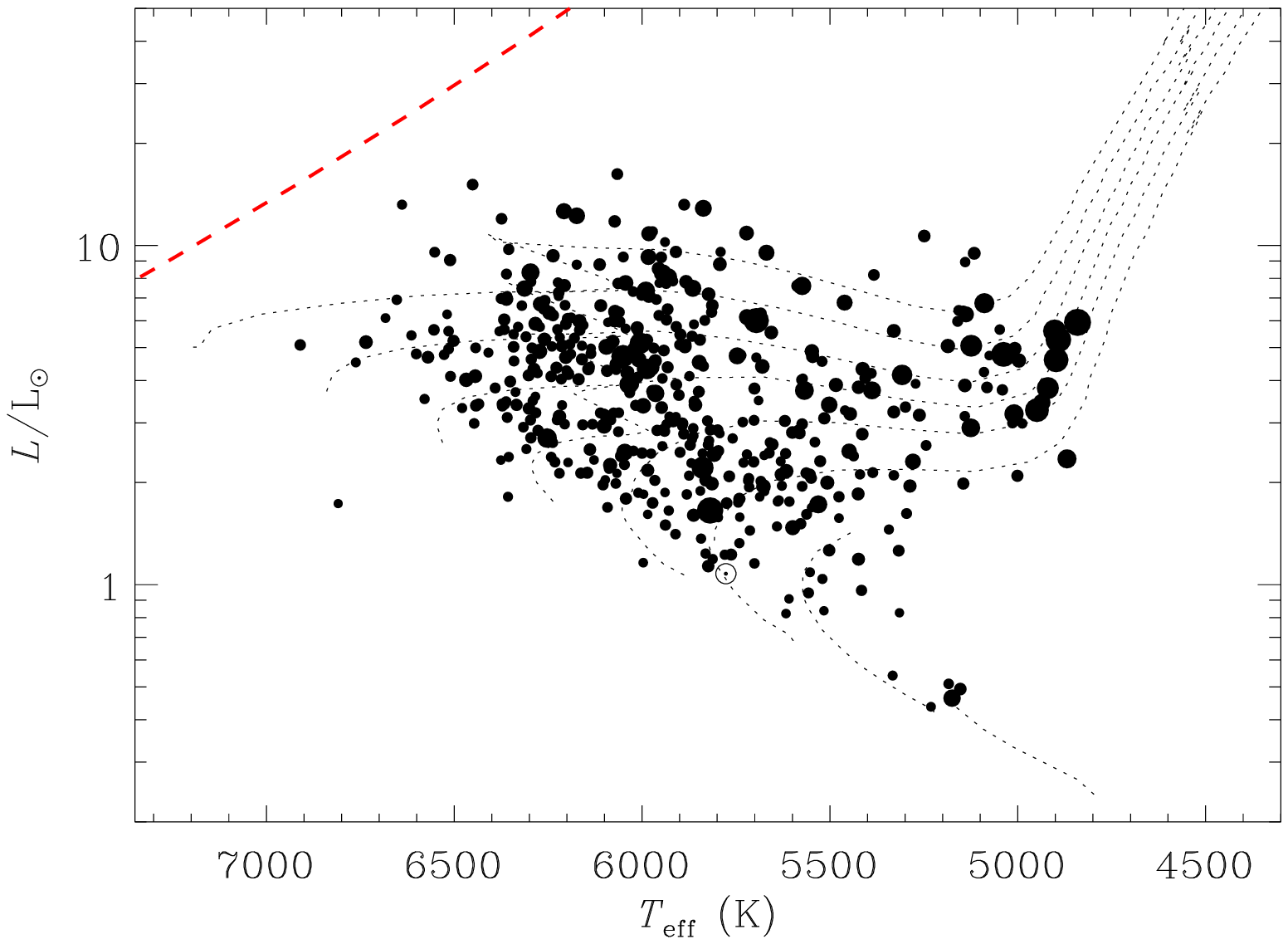}
\caption{Bottom panel: H-R diagram of stars that showed detections of
 solar-like oscillations in their \emph{Kepler} Q1, Q2 or Q3 survey
 data. Symbol size is proportional to $\rm SNR_{\rm tot}$. The
 location of the Sun is marked with the usual solar symbol. The dotted
 lines are evolutionary tracks (Padova models) for solar composition,
 computed for masses ranging from 0.7 to $1.5\,\rm M_{\odot}$ (see
 text). The dashed red line marks the approximate location of the red
 edge of the instability strip. Top left-hand panel: Predictions based
 on KIC parameters of all observed solar-type targets, but without the
 $\beta$ correction, showing stars for which $P_{\rm final} \ge
 0.9$. Top right-hand panel: Predictions after application of the
 $\beta$ correction to the scaling relation for the oscillation
 amplitudes.}
\label{fig:HR}
\end{figure*}


The bottom panel of Fig.~\ref{fig:HR} plots stars that showed
detections of solar-like oscillations with $\nu_{\rm max} \ge 350\,\rm
\mu Hz$, with luminosities estimated from the seismically determined
radii. We demanded that at least two of the asteroseismic data
analysis pipelines returned consistent results on a star, giving a
total of 499 stars with flagged detections. This is out of a total of
1750 observed stars having predicted $\nu_{\rm max} \ge 350\,\rm \mu
Hz$. The sizes of the symbols are proportional to the observed $\rm
SNR_{\rm tot}$. The dotted lines in Fig.~\ref{fig:HR} are evolutionary
tracks (Padova models; Giradi et al. 2002, 2004; Marigo et al. 2008)
computed for solar composition and masses ranging from 0.7 to
$1.5\,\rm M_{\odot}$ in steps of $0.1\,\rm M_{\odot}$. The dashed red
line marks the location of the red edge of the instability strip (for
fundamental radial-mode $\delta$-Scuti pulsations; see Houdek et
al. 1999). The Sun is plotted with its usual symbol. The lack of
detections close to the red edge, along with the amplitude ratio data
plotted in Fig.~\ref{fig:beta}, will provide important inputs for
modeling the excitation and damping of solar-like oscillations (i.e.,
the interplay between those oscillations and the convection).

The top panels in Fig.~\ref{fig:HR} show results from the predictions,
which were made using the KIC parameters of the stars, and only stars
for which $P_{\rm final} \ge 0.9$ were plotted. (Note that here the
luminosities were estimated from the KIC parameters -- using the
radius and $T_{\rm eff}$ -- whereas the luminosities in the lower
panel are the observed luminosities, from the seismically determined
radii and $T_{\rm eff}$.)  Symbol sizes are proportional to the
\emph{predicted} $\rm SNR_{\rm tot}$. The top left-hand panel shows
predictions \emph{without} the $\beta$ correction having been
applied. A total of 1163 stars are flagged with predicted
detections. This emphasizes the importance of applying the $\beta$
correction, since without it the predicted detections run all the way
to the red-edge of the instability strip, and the predicted $\rm
SNR_{\rm tot}$ reach noticably higher values across the ensemble than
do the real observations. When the $\beta$ correction is applied, we
obtain the results plotted in the top right-hand panel. There is now
much better agreement with the observations, and 765 stars are flagged
with predicted detections.

\begin{figure*}
\epsscale{1.0}
\plottwo{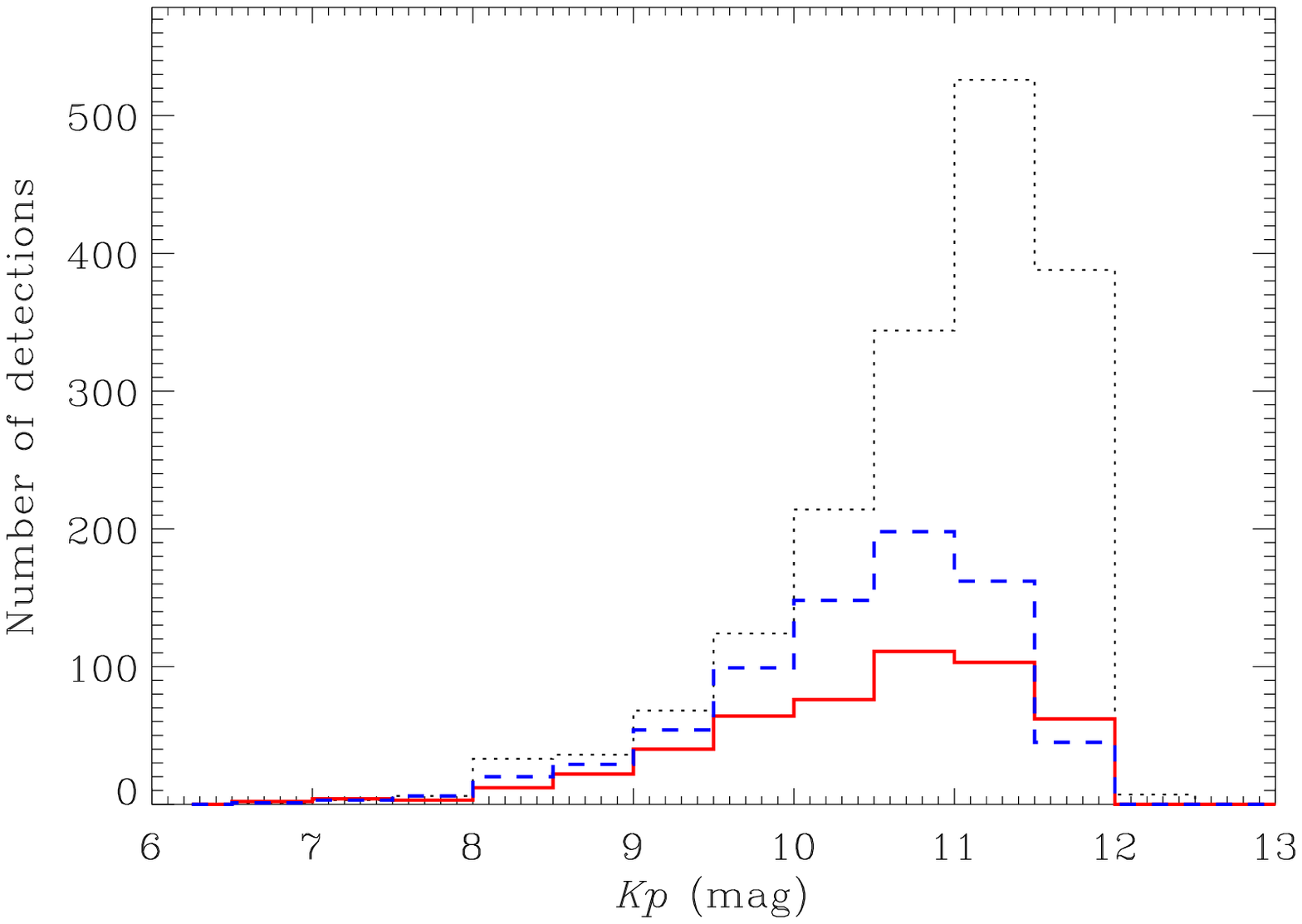}{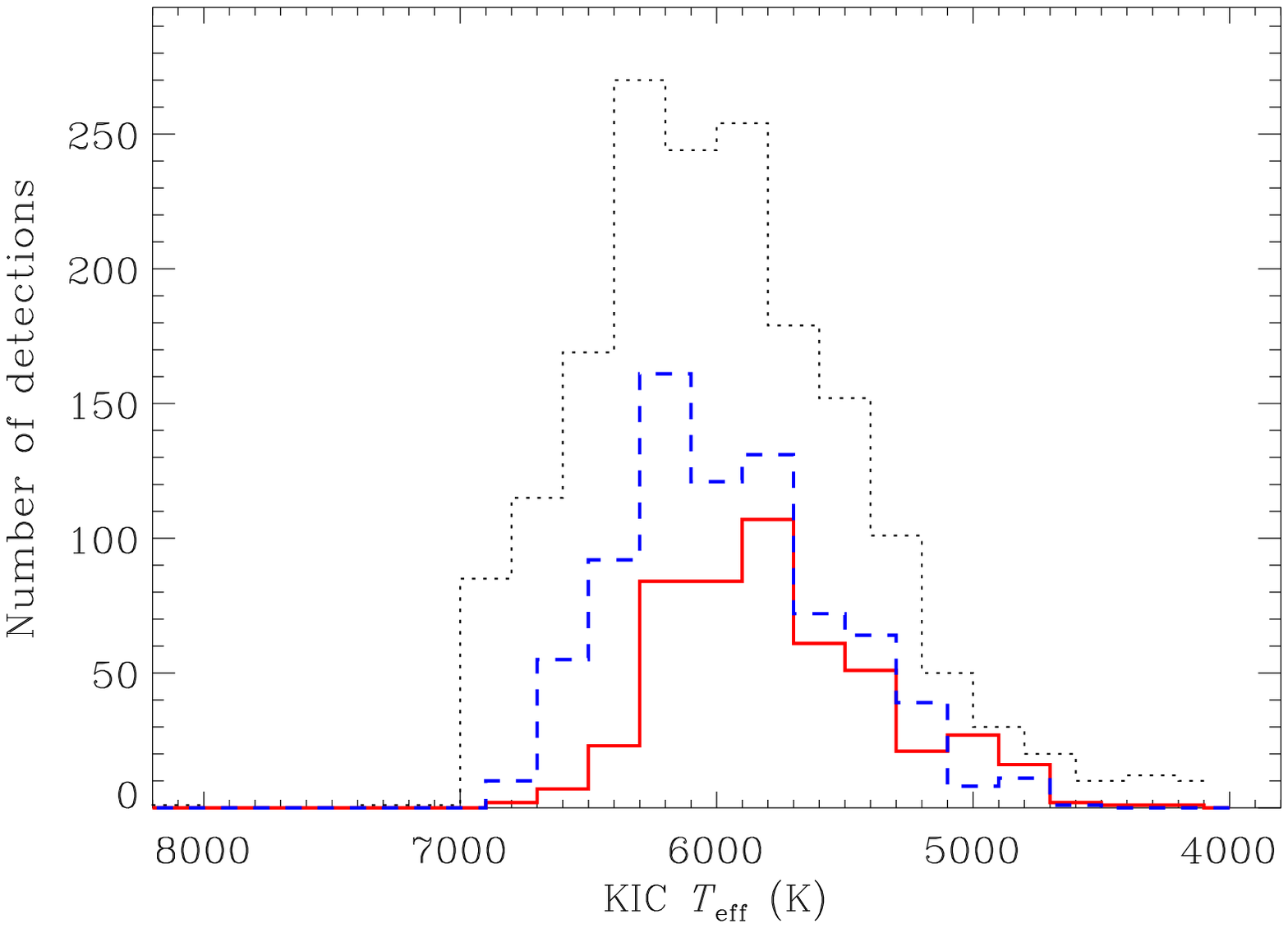}
\caption{Histograms of the observed detections (solid red lines), the
$\beta$-corrected predictions (dashed blue lines), and the total
number of observed stars having predicted $\nu_{\rm max} \ge 350\,\rm
\mu Hz$ (dotted lines), as functions of KIC apparent magnitude, $Kp$
(left-hand panel) and KIC $T_{\rm eff}$ (right-hand panel).}
\label{fig:hist}
\end{figure*}

Fig.~\ref{fig:hist} shows histograms of the observed detections (solid
red lines), the $\beta$-corrected predictions (dashed blue lines), and
the total number of observed stars having predicted $\nu_{\rm max} \ge
350\,\rm \mu Hz$ (dotted lines), as functions of KIC apparent
magnitude, $Kp$ (left-hand panel) and KIC $T_{\rm eff}$ (right-hand
panel). We see good agreement between the distributions of observed
detections and $\beta$-corrected predictions.


The predicted number of detections exceeds the observed number by
about 50\,\%. Several factors contribute to this overestimation.
First, it turns out that there are several classical oscillators in
the ensemble. The KIC parameters imply that these stars should show
detectable solar-like oscillations; however, owing to uncertainties in
the categorization it appears that they are actually too hot. Second,
there are several eclipsing binaries in the sample, which can make
automated extraction of the oscillations parameters more challenging
(due to the presence of fairly sharp features in the lightcurves).
Neither of the above categories of stars is sufficiently numerous or
troubling to account for most of the overestimation. It appears
instead that elevated levels of stellar activity might be responsible
for suppression of the mode amplitudes, and hence levels of
detectability, in a sufficient number of stars to possibly explain the
mismatch (as reported in Chaplin et al. 2011b). This may also relate
to an interesting difference between the predictions and observations
in the set of stars covering $6 \la L/{\rm L_{\odot}} \la 9$ and $5300
\la T_{\rm eff} \la 5700\,\rm K$ in the upper right-hand panel of
Fig.~\ref{fig:HR}, which are largely missing from the lower
panel. Gilliland (1985) has noted that stars in this region may show
interesting evolutionary effects in their stellar dynamos, which would
manifest themselves in the surface magnetic activity. We also add a
note of caution regarding stars in this part of the H-R diagram. They
have $\nu_{\rm max} \sim 350$ to $\sim 500\,\rm \mu Hz$. This range in
frequency coincides with an instrumental artefact in the \emph{Kepler}
data (see Garc\'ia et al. 2010), which may hamper detection of the
solar-like oscillations.

\begin{figure*}

\plottwo{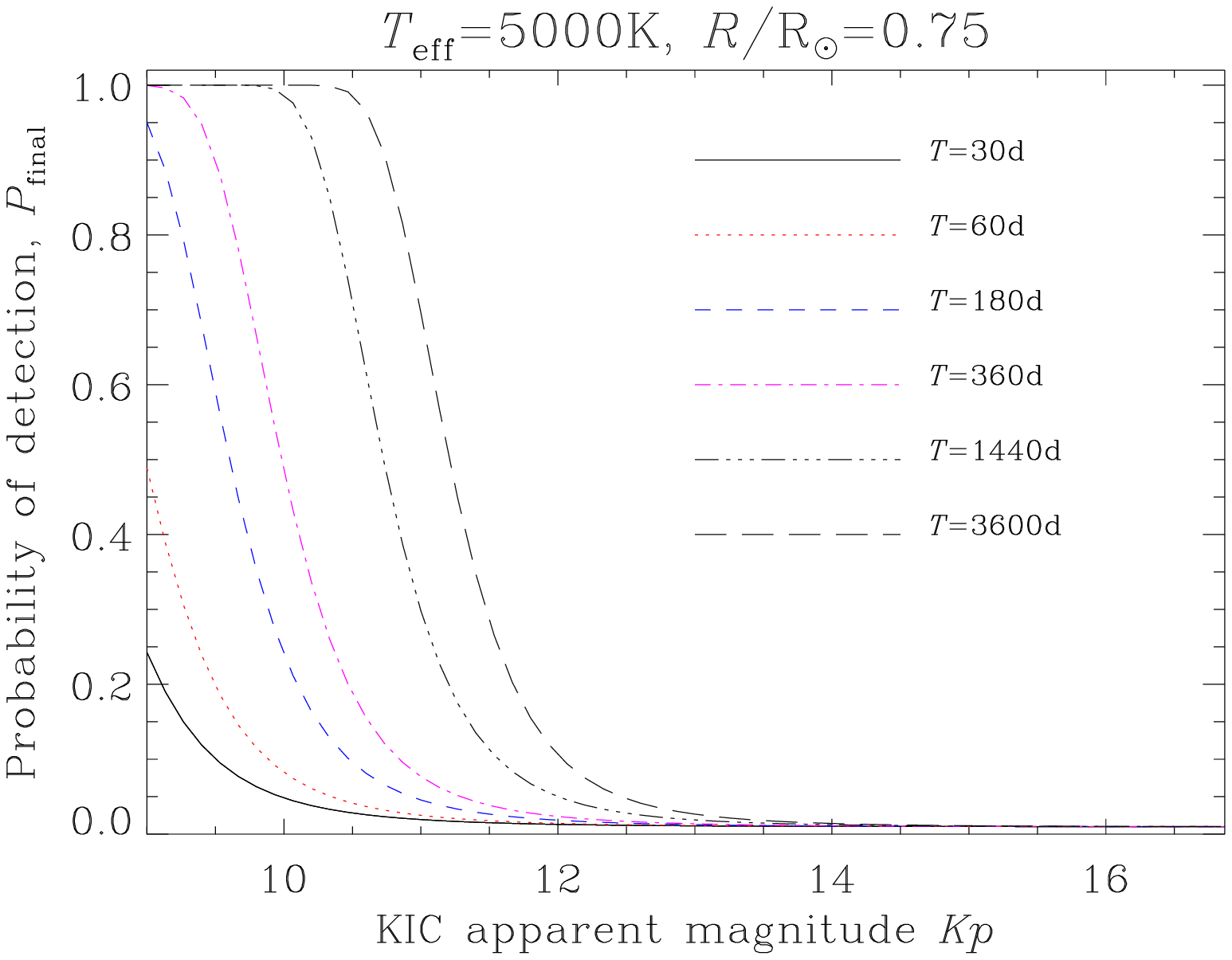}{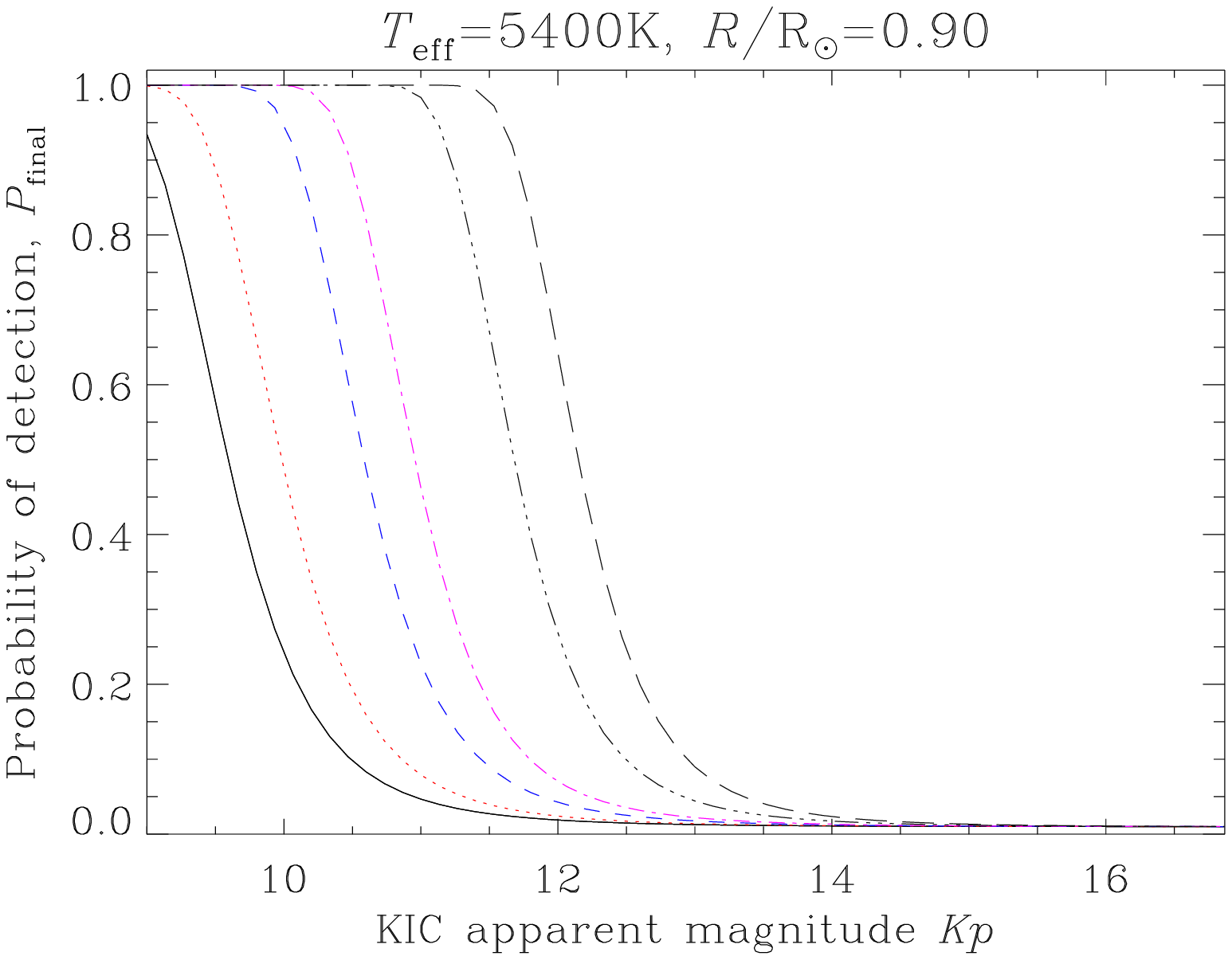}

\plottwo{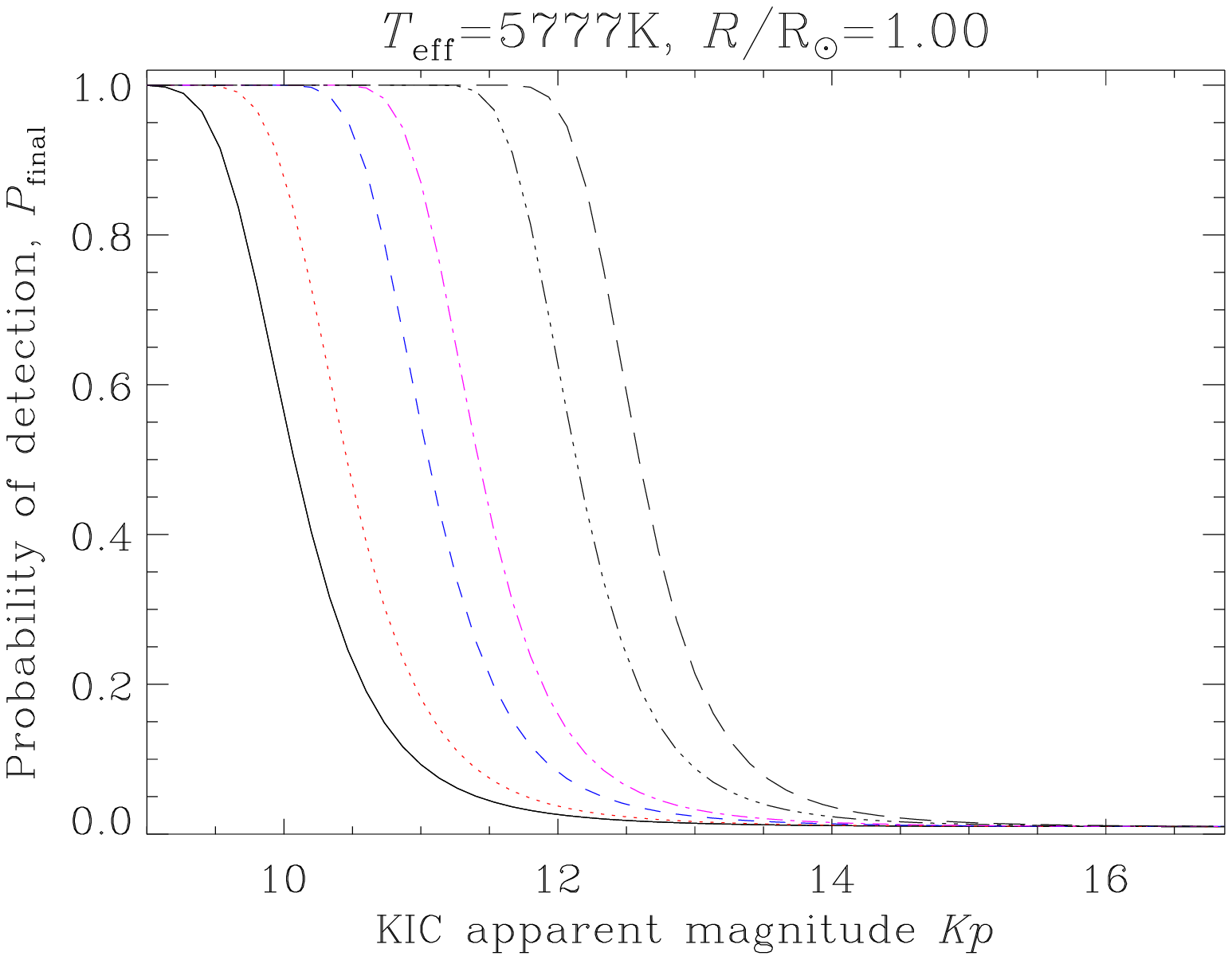}{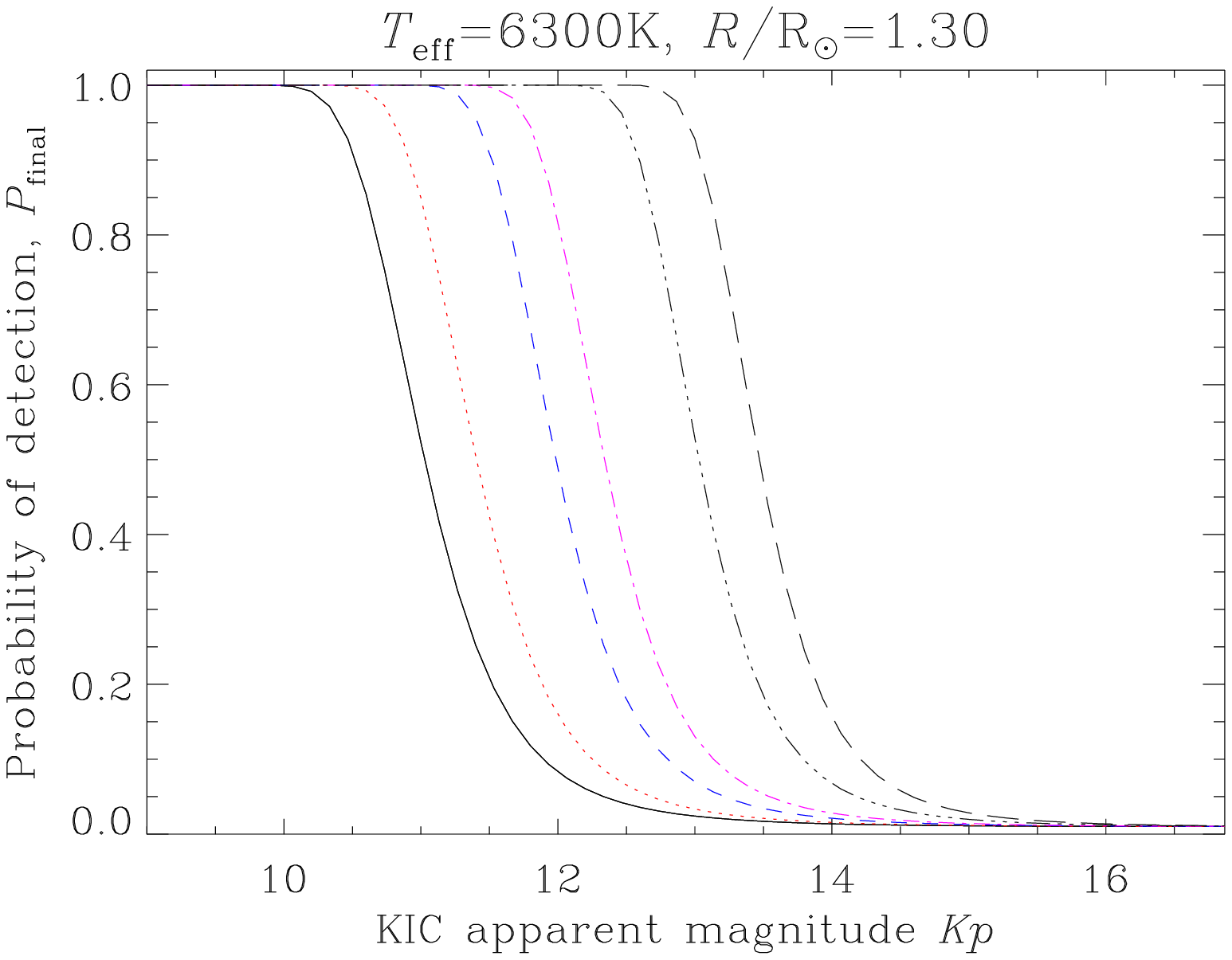}

\caption{Detection probabilities, $P_{\rm final}$, of four different
main-sequence stars for different assumed lengths of observation, $T$
(see anotation) as a function of \emph{Kepler} apparent magnitude,
$Kp$}
\label{fig:preds}
\end{figure*}

We finish by using the detection test to predict the detection
probabilities for four canonical main-sequence stars: early-type and
late-type K stars, the Sun (late G), and a mid-type F star.
Fig.~\ref{fig:preds} plots the detection probabilities, $P_{\rm
final}$, for different assumed lengths of observation, $T$ (see figure
annotation), as a function of \emph{Kepler} apparent magnitude,
$Kp$. The detection rates fall off quite sharply with $Kp$, and long
timescale observations are required in order to detect oscillations in
stars fainter than $Kp \sim 13$.


\acknowledgements Funding for this Discovery mission is provided by
NASA's Science Mission Directorate. The authors wish to thank the
entire \emph{Kepler} team, without whom these results would not be
possible.  We also thank all funding councils and agencies that have
supported the activities of KASC Working Group\,1. WJC, YE, SJH and
GAV acknowledge the support of the UK Science and Technology
Facilities Council (STFC). SH acknowledges support from the
Netherlands Organisation for Scientific Research (NWO). NCAR is
supported by the National Science Foundation.  We are also grateful
for support from the International Space Science Institute (ISSI).


\end{document}